\documentclass[%
 reprint           , 
 superscriptaddress,
 altaffilletter    ,
 showpacs          ,
 showkeys          ,
 aps               ,
 prl               ,
 amsmath, amssymb, aps,prl, floatfix,
]{revtex4-2}

\usepackage{graphicx}

\usepackage{multirow}
\usepackage{dcolumn} 

\usepackage{amsmath}
\usepackage{amssymb}
\usepackage[version=3]{mhchem}
\usepackage{siunitx}

\usepackage[hidelinks,linktocpage,colorlinks]{hyperref}
\hypersetup{
            colorlinks,
            linkcolor={red!70!black}  ,
            citecolor={green!70!black},
            urlcolor ={blue!70!black}
           }

\usepackage{comment} 
\usepackage[switch]{lineno} 
\usepackage[usenames,dvipsnames]{xcolor}
\usepackage{xspace}

\newcommand{\bb}   {\ensuremath{0\nu\beta\beta}\xspace}
\newcommand{\bbvv} {\ensuremath{2\nu\beta\beta}\xspace}

\newcommand{\CUORE}{CUORE\xspace}

\newcommand{\MC}   {MC\xspace} 

\newcommand{\tHL}  {\ensuremath{T_{1/2}^{2\nu}}\xspace}                   
\newcommand{\Nobs} {\ensuremath{N^{2\nu}_{\rm obs}}\xspace}               
\newcommand{\IA}   {\ensuremath{\mathcal{IA}_{\mathrm{^{130}Te}}\xspace}} 
\newcommand{\mTeO} {\ensuremath{m_{\rm TeO_{2}}}\xspace}                  
\newcommand{\expo} {\ensuremath{\varepsilon_{\mathrm{TeO_2}}}\xspace}     
\newcommand{\gA}   {\ensuremath{g_{\rm A}}\xspace}                        
\newcommand{\gAeff}{\ensuremath{g_{\rm A}^{\text{eff}}}\xspace}           

\begin{document}

\title{Half-life and precision shape measurement of \texorpdfstring{$\boldsymbol{\bbvv}$}{two-neutrino double beta} decay of \texorpdfstring{\ce{^{130}Te}}{Te-130}}

\author{D.~Q.~Adams}
\affiliation{Department of Physics and Astronomy, University of South Carolina, Columbia, SC 29208, USA}

\author{C.~Alduino}
\affiliation{Department of Physics and Astronomy, University of South Carolina, Columbia, SC 29208, USA}

\author{K.~Alfonso}
\affiliation{Center for Neutrino Physics, Virginia Polytechnic Institute and State University, Blacksburg, Virginia 24061, USA}

\author{F.~T.~Avignone~III}
\affiliation{Department of Physics and Astronomy, University of South Carolina, Columbia, SC 29208, USA}

\author{O.~Azzolini}
\affiliation{INFN -- Laboratori Nazionali di Legnaro, Legnaro (Padova) I-35020, Italy}

\author{G.~Bari}
\affiliation{INFN -- Sezione di Bologna, Bologna I-40127, Italy}

\author{F.~Bellini}
\affiliation{Dipartimento di Fisica, Sapienza Universit\`{a} di Roma, Roma I-00185, Italy}
\affiliation{INFN -- Sezione di Roma, Roma I-00185, Italy}

\author{G.~Benato}
\affiliation{Gran Sasso Science Institute, L'Aquila I-67100, Italy}
\affiliation{INFN -- Laboratori Nazionali del Gran Sasso, Assergi (L'Aquila) I-67100, Italy}

\author{M.~Beretta}
\affiliation{Dipartimento di Fisica, Universit\`{a} di Milano-Bicocca, Milano I-20126, Italy}
\affiliation{INFN -- Sezione di Milano, Milano I-20133, Italy}

\author{M.~Biassoni}
\affiliation{INFN -- Sezione di Milano Bicocca, Milano I-20126, Italy}

% ...
\collaboration{CUORE Collaboration}
\author{D. Castillo}
 \affiliation{Departament de F\'{i}sica Qu\`{a}ntica i Astrof\'{i}sica, Universitat de Barcelona , 08028 Barcelona, Spain}
 \affiliation{Institut de Ci\`{e}ncies del Cosmos, Universitat de Barcelona, 08028 Barcelona, Spain}

\author{J.\ Kotila}
 \affiliation{Finnish Institute for Educational Research, University of Jyv\"askyl\"a, P.O. Box 35, FI-40014, Jyv\"askyl\"a, Finland}
 \affiliation{International Centre for Advanced Training and Research in Physics (CIFRA), P.O. Box MG12, 077125 Bucharest-Magurele, Romania}
 
\author{J. Men\'{e}ndez}
 \affiliation{Departament de F\'{i}sica Qu\`{a}ntica i Astrof\'{i}sica, Universitat de Barcelona , 08028 Barcelona, Spain}
 \affiliation{Institut de Ci\`{e}ncies del Cosmos, Universitat de Barcelona, 08028 Barcelona, Spain}

\author{O.\ Ni\c{t}escu}
 \affiliation{International Centre for Advanced Training and Research in Physics (CIFRA), P.O. Box MG12, 077125 Bucharest-Magurele, Romania}
 \affiliation{“Horia Hulubei” National Institute of Physics and Nuclear Engineering, 30 Reactorului, P.O. Box MG6, 077125 Bucharest-M\u{a}gurele, Romania}

\author{F.\ \v{S}imkovic}
 \affiliation{Faculty of Mathematics, Physics and Informatics, Comenius University in Bratislava, 842 48 Bratislava, Slovakia}
 \affiliation{Institute of Experimental and Applied Physics, Czech Technical University in Prague, 110 00 Prague, Czech Republic}

\date{\today}

\begin{abstract}
 We present a new measurement of the \bbvv half-life of \ce{^{130}Te} (\tHL) using the first complete model of the \CUORE data, based on 1038~kg\,yr of collected exposure.
 Thanks to optimized data selection, we achieve a factor of two improvement in precision, obtaining $\tHL~=~\left( 9.32\,^{+0.05}_{-0.04}\, \text{stat.}\,^{+0.07}_{-0.07}\,\text{syst.} \right)~\times~10^{20}~\text{yr}$. 
 The signal-to-background ratio is increased by $70\%$ compared to our previous results, enabling the first application of the improved \bbvv formalism to \ce{^{130}Te}.
 Within this framework, we determine a credibility interval for the effective axial coupling in the nuclear medium as a function of nuclear matrix elements. We also extract values for the higher-order nuclear matrix element ratios: second-to-first and third-to-first.
 The second-to-first ratio agrees with nuclear model predictions, while the third-to-first ratio deviates from theoretical expectations.
 These findings provide essential tests of nuclear models and key inputs for future \bb searches.
\end{abstract}

\maketitle

 \noindent Two-neutrino double beta decay (\bbvv) is a rare nuclear process where two neutrons in a nucleus transform into two protons, emitting two electrons and two electron antineutrinos~\cite{Saakyan:2013yna}. 
 The study of \bbvv has significant implications in nuclear physics, as a precise measurement of the transition \mbox{half-life} and spectral shape can help test and refine \mbox{nuclear-structure} models for heavy nuclei.
 At the same time, \bbvv is a background for neutrinoless double beta decay (\bb), a hypothetical process that carries major implications for physics, including understanding whether the lepton number is a conserved quantity and whether the neutrino is a Majorana particle~\cite{Agostini:2022zub}.
 Nowadays, the major results on \bbvv are indeed coming from experiments whose primary goal is the search for \bb~\cite{EXO-200:2013xfn,KamLAND-Zen:2019imh,CUORE:2020bok,GERDA:2023wbr,CUPID:2023wyy,CUPID-Mo:2023lru}.

 The latest measurement of the \bbvv \mbox{half-life} of \ce{^{130}Te} has been performed by the Cryogenic Underground Observatory for Rare Events (\CUORE, \cite{CUORE:2021mvw}), an experiment located underground at the Laboratori Nazionali del Gran Sasso in Italy.
 The \CUORE detector consists of $988$ \ce{TeO_2} cryogenic calorimeters arranged in $19$ towers of $13$ \mbox{four-crystal} floors, and each crystal is read out as an individual channel.
 The array has a total mass of $742$~kg. 
 It is housed inside a large custom cryostat cooled to about $15$~mK by a \ce{^3He}/\ce{^4He} dilution refrigerator~\cite{Alduino:2019xia}, which provides a \mbox{low-noise} and low-radioactivity environment.
 
 In this Letter, we present a refined study of the \bbvv by \CUORE, along with the first-ever application to \ce{^{130}Te} of the improved formalism for the description of the \bbvv process~\cite{Simkovic:2018rdz}. 
 As well as increased statistics, this study benefits from the first full model of the \CUORE background~\cite{CUORE:2024fak}, validated by over $1$~t\,yr of data across the full energy range and improved treatment of multi-crystal events, i.\,e.\ simultaneous energy depositions sharing a common particle-physics origin. 
 Although the primary objectives of the background model are the reconstruction of the data as a linear combination of different components and the determination of the corresponding activities, the same infrastructure can be utilized for a precise determination of \bbvv \mbox{half-life} of \ce{^{130}Te}, referred to as \tHL.

 To maximize our ability to precisely assess the shape of \bbvv, we adopt optimized data selection criteria, prioritizing a higher \mbox{signal-to-background} ratio and goodness of fit over a larger exposure.
 In particular, in Ref.~\cite{CUORE:2024fak}, all \CUORE crystals have been included in the analysis, and a rigorous evaluation of systematic uncertainties has been performed to ensure that their potential impact is addressed conservatively.
 Conversely, here we consider only the crystals of the innermost towers ($7$ out of $19$) to profit from the detector self-shielding against the external radiation, that is more difficult to model; we restrict the energy range to remove the part below $250$~keV, where the data reconstruction is weaker; and we reduce the width of the binning to perform a more detailed description of the spectrum.
 Overall, this translates into a cumulative increase of the \bbvv events over the background of approximately $70\%$ and a threefold improvement of the goodness of fit, mainly coming from the geometric crystal selection.

 The data used for the analysis amounts to $15$ datasets, where a dataset denotes a collection of physics runs with steady operating conditions flanked by calibration runs.
 Since the status of the detector directly affects both the \mbox{data-taking} (dead channels, unusable time intervals, etc.) and the analysis (energy thresholds, data quality cuts, etc.), each dataset undergoes the complete \mbox{data-processing} chain~\cite{TMPCUOREanalysis}, and we compute the efficiencies for each \mbox{channel-dataset} pair. 
 At the same time, individual \mbox{channel-dataset} pairs are excluded when it is not possible for them to go through the whole analysis chain, due to e.\,g.\ low-quality calibration or insufficient statistics to determine the peak lineshape.
 In order to simulate a realistic \mbox{\CUORE-like} set of data, $2\nu\beta\beta$ and background events are generated across all the channels and datasets, and all the data-taking and analysis efficiencies are included within the Monte Carlo (\MC) processing. 
 
 \begin{figure}[tb]
   \centering
   \includegraphics[width=.99\columnwidth]{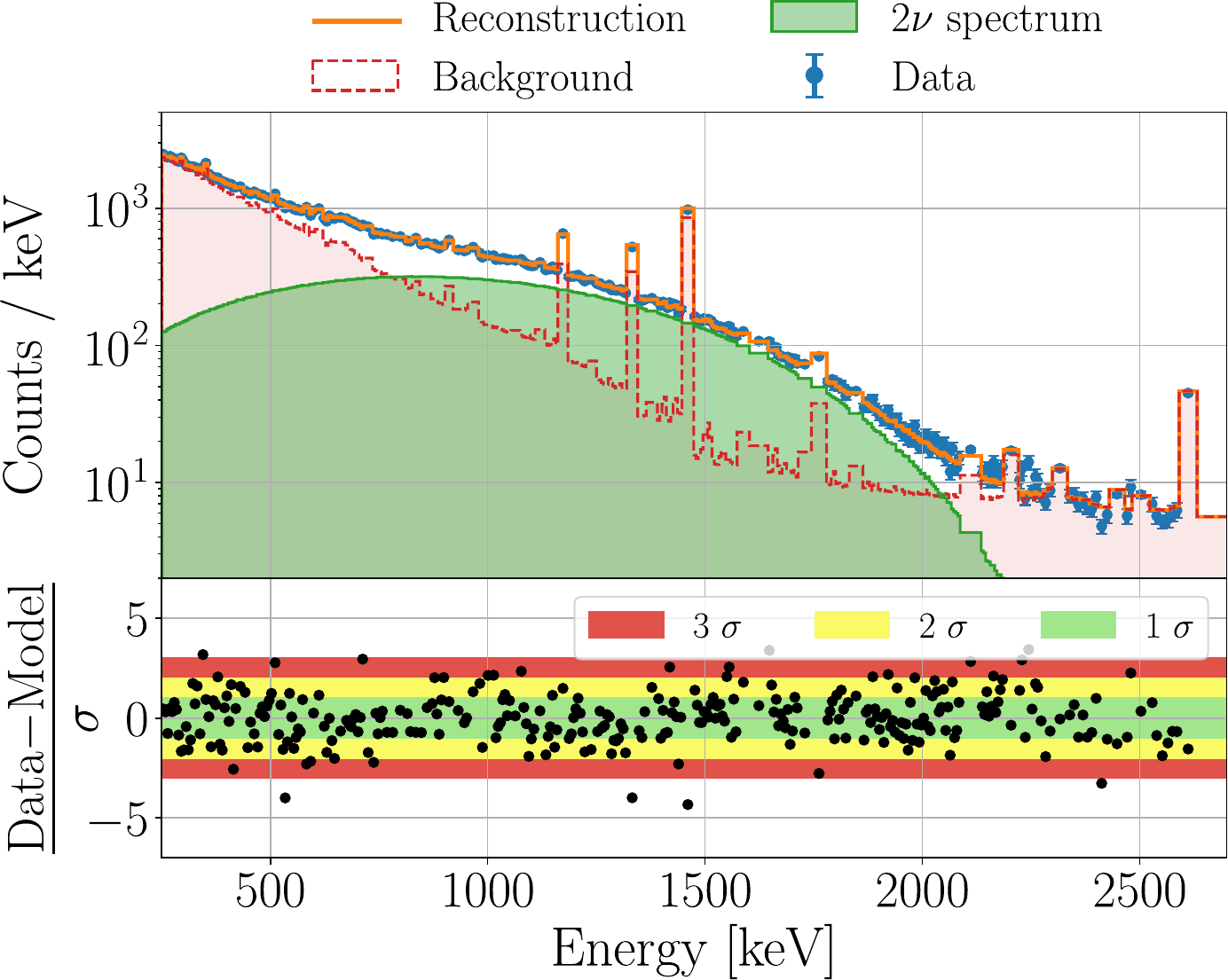}
   \caption{Best fit to the single-crystal-event energy spectrum obtained by \CUORE after $1038.4$~kg\,yr of accumulated exposure. The reconstructed histogram (orange) is overlayed to the experimental data (blue). The individual \bbvv contribution within the SSD model is represented by the green area, while the overall background is shown in red.
   The lower panel shows the difference between the data and the best-fit model normalized to the statistical uncertainty overlaid to the $1\sigma$, $2\sigma$ and $3\sigma$ bands.}
   \label{fig:2nuReconstruction}
 \end{figure}

 We selected the Single State Dominance (SSD,~\cite{Abad:1984XXX}) as the reference model for the \bbvv decay.
 The SSD model assumes that the lowest $1^{+}$ state of the intermediate nucleus provides the leading contribution to both spectral shape and decay \mbox{half-life}.
 As an alternative, we also examined the Higher State Dominance (HSD), which instead predicts a significant contribution to the intermediate nucleus from states higher than the lowest $1^+$. 
 We obtain a reduced chi-square $\chi_{\rm red}^{2}=1.47$ ($1339$ d.\,o.\,f.) with the former model, while the latter yields a $\chi_{\rm red}^{2}=1.51$.
 Figure~\ref{fig:2nuReconstruction} shows the optimized fit within the SSD model to the \CUORE single-crystal-event energy spectrum zoomed-in over the region 250--2700~keV, thus highlighting the contribution from the \bbvv decay. 
  
 We determine \tHL from the expression:
 \begin{equation}
  \label{eq:2nu}
  \tHL = \ln 2\ \frac{N_{\rm A}~ \expo~ \IA}{\mTeO}\ \frac{\epsilon}{\Nobs},
 \end{equation}
 where $N_{\rm A}$ is the Avogadro number, \expo is the collected \ce{TeO_2} exposure ($1038.4$~kg\,yr), \IA~is the isotopic abundance of \ce{^{130}Te} ($0.34167 \pm 0.00002$,~\cite{Fehr200483}), \mTeO is the molar mass of \ce{TeO_{2}} (159.6~g\,mol$^{-1}$), $\epsilon$ is the probability to observe a \bbvv event in the analysis range and $N_{\rm obs}^{2\nu}$ is the number of observed \bbvv decays as determined by the fit.
 Since we adopted a Bayesian approach, the statistical uncertainty associated with the measurement is entirely estimated from the narrowest $68\%$ interval around the mode of the \bbvv normalization posterior, and it is $O(0.5\%)$.
 The contributions from \IA, \mTeO, \expo, and $\epsilon$ are negligible. 
 Of these, $\epsilon$ is the dominant one at $O(0.01\%)$.

 We consider multiple sources of systematic uncertainties that could impact the determination of \tHL.
 First, we benchmarked our model against the variation of the energy range used for the fit.
 In particular, the low-energy region of the spectrum
 is populated by multiple featureless background components, making the reconstruction challenging and susceptible to biases due to unidentified or mismodeled background sources.
 To quantify the impact of this, we repeated the fit by varying the low-energy threshold within 200--300~keV in steps of $25$~keV.
 
 We also tried different sizes for the binning to assess whether a specific choice could either reveal or hide significant spectral features. 
 We included all the possible combinations of bin widths in $[3, 5, 7]$~keV and $[5, 10, 20]$~keV for \mbox{single-crystal} and \mbox{multi-crystal} events, respectively.
 
 In the background model, we assume that contaminants remain constant over time and are uniformly distributed within the simulated volumes.
 To test the validity of these assumptions, we performed different fits by organizing the \CUORE datasets and channels into several subsets.
 We defined different groups of datasets ordering them both chronologically and randomly. 
 Since \CUORE has shown to be sensitive to seasonal variations of environmental noise, which affects parameters such as the energy resolution and the trigger threshold~\cite{TMPCUOREanalysis}, the datasets have also been organized based on the data-taking period.
 To test the assumption of uniform-contaminant distribution, we performed fits on subsets of the detector, dividing it in even and odd channels, upper and lower floors and innermost versus outermost columns.

 Furthermore, since the Bremsstrahlung \mbox{cross-section} has been found to be a limiting factor for the \MC accuracy~\cite{Batic:2013sta,Pandola:2014uea}, we generated simulations of the \bbvv decay where the cross-section value is varied by $\pm 10\%$.
  
 \begin{table}[tb]
  \centering
  \caption{Sources of systematic uncertainty affecting the \bbvv half-life. 
  The total value associated to the measurement is computed as the sum in quadrature of the individual contributions. \\}
  \setlength{\tabcolsep}{8pt}
  \begin{tabular}{lc}
   \toprule \\[-8pt]
   Source          &Uncertainty [\%]       \\[2pt]  
   \hline   \\[-7pt]
   Energy threshold & $^{+0.146}_{-0}$     \\[5pt]
   Binning          & $^{+0.160}_{-0.068}$ \\[5pt]
   Dataset          & $^{+0.594}_{-0.594}$ \\[5pt]
   Geometry         & $^{+0.350}_{-0.350}$ \\[5pt]
   Bremsstrahlung   & $^{+0.160}_{-0.162}$ \\[5pt]
   \hline   \\[-7pt]
   Total            & $^{+0.740}_{-0.711}$ \\[3pt]
   \toprule
  \label{tab:2nuSystematics}
  \end{tabular}
 \end{table}
 
 To validate the robustness of the fit, we implemented multiple consistency checks, although these were not included in the computation of the systematic uncertainties.
 The fit was repeated assigning flat uninformative priors (except for the muon component, whose prior was independently fixed using multi-crystal \CUORE data with simultaneous depositions in several crystals~\cite{CUORE:2024fak}) or excluding background sources compatible with zero activity.
 The results showed excellent agreement and were consistent with the findings of the background model, thus demonstrating that the optimization steps did not influence the determination of \tHL.

 The possible presence of \ce{^{90}Sr}, a pure \mbox{$\beta$-emitter} produced during nuclear fallout, can impact the measurement of \tHL due to its strong \mbox{anti-correlation} with the \bbvv. 
 In this study, we found the activity of \ce{^{90}Sr} to be negligible by constraining it to that of \ce{^{137}Cs}, a nuclide produced alongside \ce{^{90}Sr} in a known ratio~\cite{UNSCEAR2000,Igarashi:2003XXX}. 
 The activity of \ce{^{137}Cs} was estimated by its characteristic $\gamma$-ray peak at $661.7$~keV.

 Table~\ref{tab:2nuSystematics} summarizes the impact of the different sources of systematics.
 We assumed all these effects to be independent and computed the overall uncertainty by summing in quadrature all the contributions.
 
 \begin{figure}[tb]
  \centering
  \includegraphics[width=.99\columnwidth]{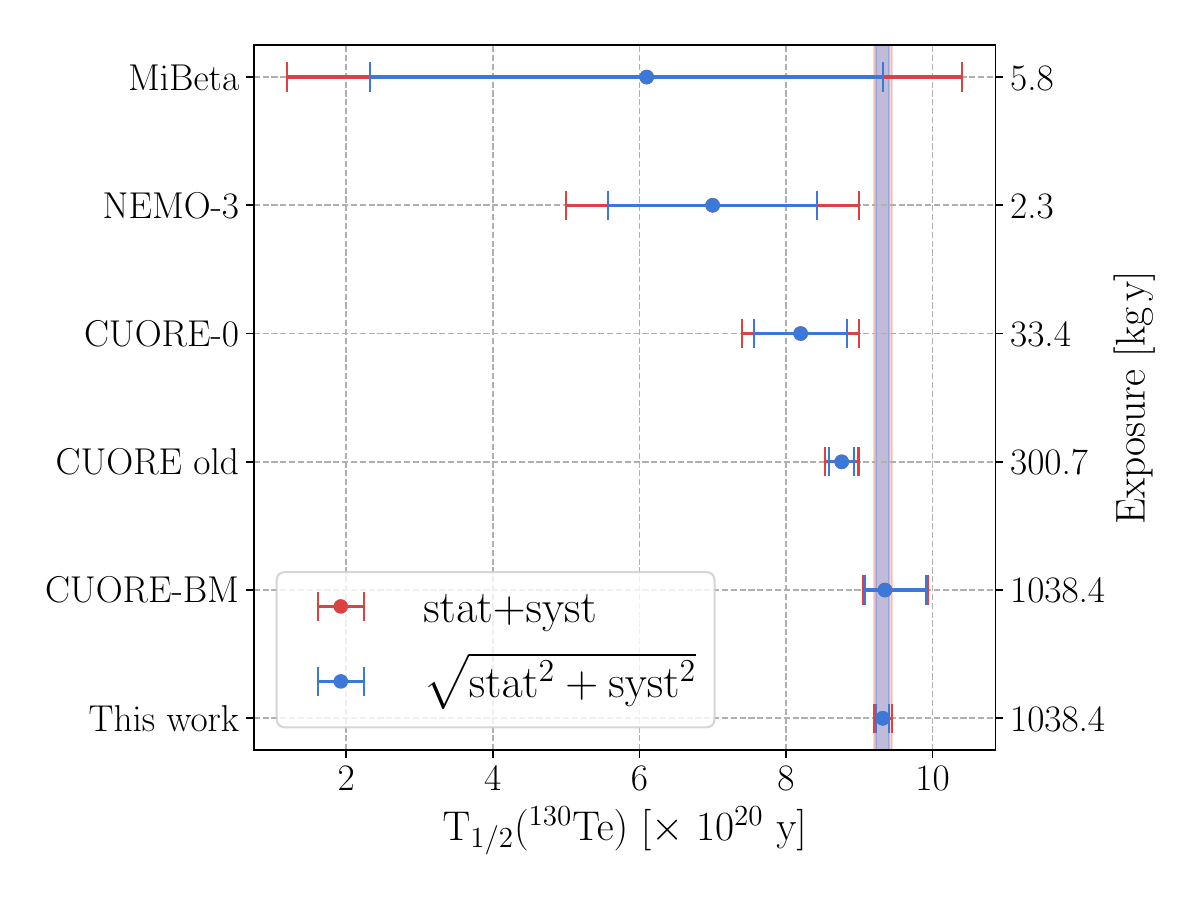}
  \caption{Measurements of the \ce{^{130}Te} \mbox{half-life} with the corresponding exposure~\cite{Arnaboldi:2002te}~\cite{Arnold:2011gq}~\cite{Alduino:2016vtd}~\cite{CUORE:2020bok}~\cite{CUORE:2024fak}.
    The error bands represent two types of uncertainty combinations: sum in quadrature (blue) and linear sum (red) of the $68\%$ statistical and systematic uncertainties. 
    The trend towards increasing values of \tHL is likely due to the improved modeling of the background over time, where the introduction of additional contributions reduces the normalization of \bbvv.}
  \label{fig:halfLifeHistory}
 \end{figure}
 
 The final result on the \bbvv half-life is:
 \begin{equation}
 \label{eq:HalfLifeFinal}
  \tHL = \left( 9.32\,^{+0.05}_{-0.04}\, \text{stat.}\,^{+0.07}_{-0.07}\,\text{syst.} \right) \times 10^{20} ~\text{yr}.
 \end{equation}
 This measurement represents a twofold improvement in precision over the previous result by CUORE~\cite{CUORE:2020bok}, with a significant reduction in the systematic uncertainties, mainly attributed to the upgraded data model and the \mbox{data-selection} criteria specifically optimized for this study.
 As a comparison, a summary of the past \bbvv \mbox{half-life} measurements has been reported in Fig.~\ref{fig:halfLifeHistory}: a trend towards increasing values of \tHL is clearly visible. 
 This feature is not associated with a physical effect but rather with the analysis. 
 As statistics increase and data models become more complex, previously neglected or unknown sub-dominant background components are included, in turn reducing the relative normalization of the \bbvv contribution.
 Such an effect is more pronounced in cases where the \mbox{signal-to-background} ratio is poor and becomes less relevant as the data are better understood,
 leading to an asymptotic-like behavior. 

 The SSD model is highly effective as it provides an excellent fit to experimental data.
 However, within this framework, the \bbvv shape is fixed a priori, and it is not possible to extract sufficient information from the \bbvv fit to support a data-driven refinement of nuclear models.
 A novel approach in this regard is represented by the so-called \emph{improved formalism}~\cite{Simkovic:2018rdz,Nitescu:2021pdq}.
 The improved formalism foresees a Taylor expansion over the lepton energies, which enables the calculation of \mbox{sub-leading} nuclear matrix elements (NMEs). 
 Spectral shapes and relative strengths of the Taylor-expanded terms offer valuable constraints on intermediate states and on the effective value of the axial coupling constant in the nuclear medium \gAeff.
 The expression for \tHL within this model takes the form:
 \begin{equation}    
  \begin{split}
  \label{eq:masterFormula}
   \left[ \tHL \right] &^{-1} = \left( \gAeff \right)^4 \left| M^{2\nu}_{\rm{GT}-1} \right|^2 \biggl\{ G_0+\xi_{31} G_2 \\
                       & + \frac{1}{3}\left(\xi_{31}\right)^2 G_{22} + \left[\frac{1}{3}\left(\xi_{31}\right)^2 + \xi_{51}\right] G_4 \\
                       & + \frac{1}{3} \xi_{31} \xi_{51} G_{42}+\frac{2}{3} \xi_{31} \xi_{51} G_6\biggr\},
    \end{split}
 \end{equation} 
 where GT stands for Gamow-Teller transition, $G_0$, $G_2$, $G_{22}$, $G_4$, $G_{42}$ and $G_6$ are the phase space factors (PSFs), $M^{2\nu}_{\rm GT-1}$, $M^{2\nu}_{\rm GT-3}$, and $M^{2\nu}_{\rm GT-5}$ are the NME expansion terms, and we define the ratios $\xi_{31} \equiv M_{\rm{GT}-3}^{2\nu} / M_{\rm{GT}-1}^{2\nu}$ and $\xi_{51} \equiv M_{\rm{GT}-5}^{2\nu} / M_{\rm{GT}-1}^{2\nu}$.
 $\xi_{31}$ and $\xi_{51}$ provide complementary insights into the decay, in particular regarding the Gamow-Teller strength and structure of the intermediate nucleus states, as $M^{2\nu}_{\rm GT-1}$ is sensitive to contributions from \mbox{high-lying} states in the intermediate odd-odd nucleus, while $M^{2\nu}_{\rm GT-3}$ and $M^{2\nu}_{\rm GT-5}$ are primarily determined by the lowest-energy ones.
 Concurrently, the study of $\xi_{31}$ and $\xi_{51}$ also probes the HSD model, since this requires $\xi_{31} = \xi_{51}=0$.
 
 By inverting Eq.~\eqref{eq:masterFormula}, we can express \gAeff as a function of \tHL, the PSFs, and $M_{\rm GT-3}^{2\nu}$.
 The set of energy spectra from the Taylor expansion terms can be fitted to the \bbvv signal in order to extract both leading and \mbox{higher-order} contributions.
 The result can then be compared with the predictions of various nuclear models.
 
 In order to assess the accuracy and validity of theoretical approximations over a broad range of atomic numbers, it is necessary to test the nuclear model descriptions across different isotopes.
 Experimental investigations have already been carried out on \ce{^{136}Xe}~\cite{KamLAND-Zen:2019imh} and on \ce{^{100}Mo}~\cite{CUPID-Mo:2023lru}.
 We now present the first-ever shape studies on the \bbvv of \ce{^{130}Te}.
 It is worth noting that, the lower signal-to-background ratio in CUORE--approximately 0.5 compared to values $\mathcal{O}(10)$ in the other cases–-is compensated by the advanced background model and the sizable collected statistics and thus provide similar level of sensitivity to the $2\nu\beta\beta$ shape.
 
 We perform the fit using the same data selection and technique presented before, but treating the \bbvv contribution as a linear combination of the improved-formalism electron spectra. 
 Each \bbvv spectrum template is weighted for the PSFs, fixed in the likelihood as they can be precisely calculated, and for the normalization factors, to which uninformative flat priors are assigned.
 The PSFs are computed using the screened exact \mbox{finite-size} Coulomb wave function~\cite{Nitescu:2024tvj}.
 The NME templates are calculated by using the proton-neutron quasi-particle random-phase approximation (pnQRPA, \cite{Simkovic:2013qiy}).
 In addition, we included radiative and atomic exchange corrections for the emitted electrons, which affect both the PSFs and the spectral shape. 
 Even though the largest impact of these corrections occurs below the fit threshold, they also introduce a non-negligible measurable shift for the maximum of the energy spectrum~\cite{Nitescu:2024tvj}.
 
 The overall fit returns a reduced chi-square $\chi^{2}_{\rm red}=1.48$ ($1337$ d.\,o.\,f.), and the \bbvv contribution is compatible with the SSD shape within $1\sigma$ across the entire energy range. 
 This outcome supports the suitability of the SSD model for \ce{^{130}Te} and slightly favors it, likely due to the lower number of parameters and the weaker correlations.
 The \mbox{half-life} we obtain with the improved formalism is consistent within $1\sigma$ with the value in Eq.~\eqref{eq:HalfLifeFinal}.
 The discrepancy could be treated as a systematic uncertainty on the model, in which case we estimate an additional contribution of $-0.53\%$.

\begin{figure}[tb]
  \centering
  \includegraphics[width=.99\columnwidth]{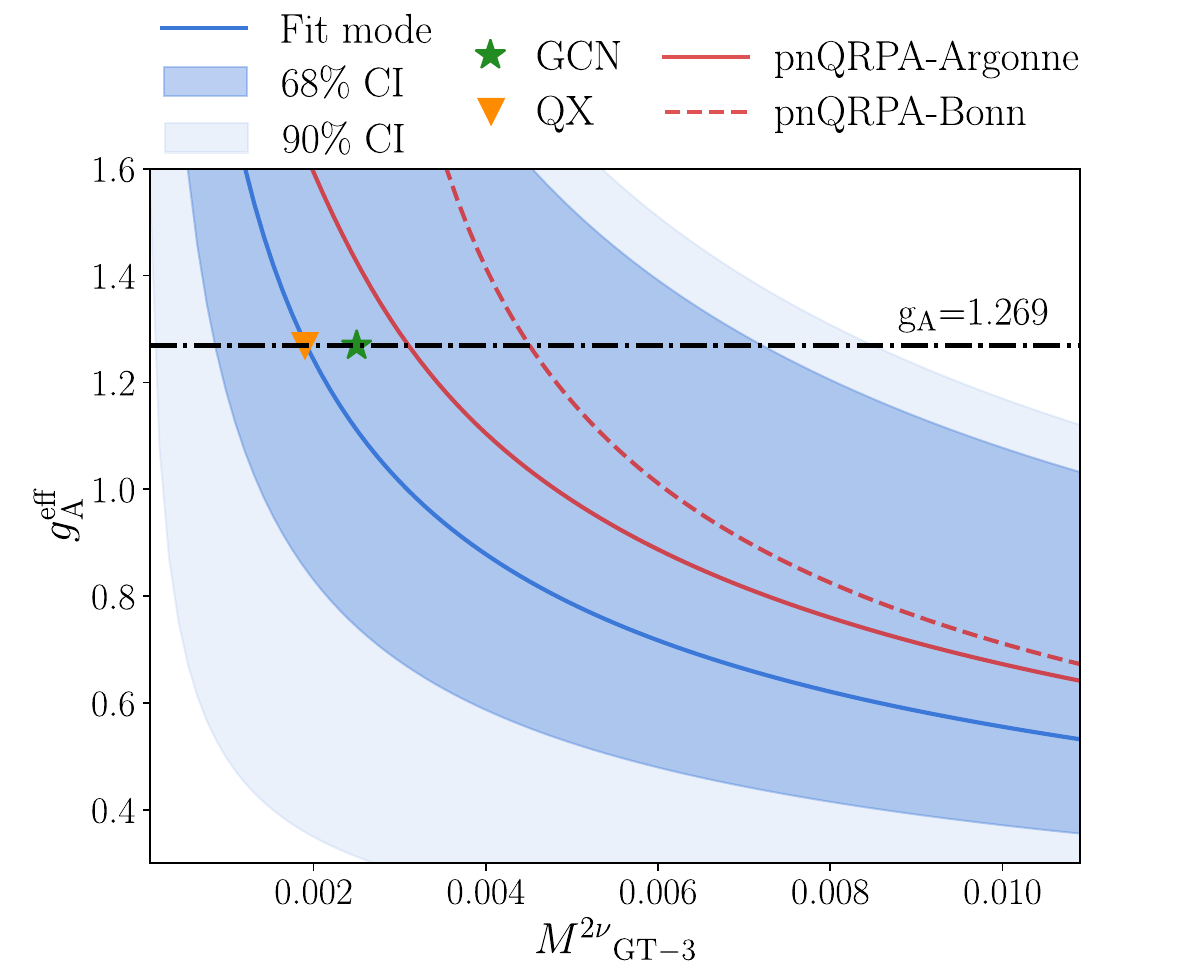}
  \caption{Effective value of the axial coupling constant \gAeff as a function of $M_{\rm GT-3}^{2\nu}$.
           The experimental fit mode is represented by the blue line and the corresponding confidence intervals by the dark blue ($68\%$ CI) and light blue ($90\%$ CI) bands.
           The theoretical predictions from the shell models are identified by the green and orange markers~\cite{Qi:2012zg,Caurier:2007wq}, while those from the pnQRPA model are indicated by the red lines, solid and dashed corresponding to two different two different kinds of short-range correlations~\cite{Simkovic:2013qiy}.}
  \label{fig:gA}
 \end{figure}

 We find $\xi_{31}$ consistent with $0$, namely $\xi_{31}=0.01_{-0.01}^{+0.31}$, and we can quote a $90\%$ credibility interval (CI) limit at $0.55$.
 In contrast, $\xi_{51} = 1.46_{-0.62}^{+0.33}$ exhibits a non-zero mode, meaning that the contribution due to the third-order NME is dominant over the one of the second-order NME (recall the definitions in Eq.~\eqref{eq:masterFormula}).
 The two parameters show an extremely high \mbox{anti-correlation} with $\rho=-0.78$, similarly to what had been previously found in Ref.~\cite{CUPID-Mo:2023lru}, where the \mbox{anti-correlation} effects were addressed by constraining the ratio $r=\xi_{51}/\xi_{31}$.
 We choose to leave $r$ free to vary, therefore showing results in the most \mbox{model-independent} way possible,
 even though uncertainties and degeneracies among the \bbvv terms lead to strong correlations.
 By applying Eq.~\eqref{eq:masterFormula} for each Markov-chain \MC step of the fit and for $M^{2\nu}_{\rm GT-3}$ in the range $0$--$0.01$, we obtain a posterior distribution for \gAeff, from which we extract $68\%$ CI and $90\%$ CI.
 The overall result is illustrated in Fig.~\ref{fig:gA}, together with predictions from two nuclear models and the \gA free-neutron value of 1.269~\cite{Simkovic:2018rdz}.
 The interacting shell model (ISM~\cite{Caurier:2004gf}) comprises two different interactions, namely QX~\cite{Qi:2012zg} and GCN5082~\cite{Caurier:2007wq}, whose \gA quenching is included in the calculation of the NMEs, and it is fixed to $0.76$ and $0.48$, respectively~\cite{Jokiniemi:2022ayc}.
 
 Figure~\ref{fig:gA} shows excellent agreement between the experimental findings and the theoretical expectations. Our value is consistent within $1\sigma$ with the theoretical predictions: $\xi_{31}$(pnQRPA) $\sim 0.2$ and $\xi_{31}$(ISM) $\sim 0.12$.
 We foresee a non-zero quenching for the axial coupling for most of the $M_{\rm GT-3}^{2\nu}$ values, although the large uncertainty does not rule out the free \gA.
 In contrast, our determination of $\xi_{51}$ significantly exceeds the theoretical values, which are approximately $0.05$ and $0.028$ for pnQRPA and ISM, respectively.
 Similar considerations apply when comparing our results with SSD predictions, that are $\xi_{31}=0.35$ and $\xi_{51}=0.123$, respectively.
 
 Values of $\xi_{51}$ lower than $1$ lie within the $68\%$ CI. However, a mode close to $1.5$ raises concerns about the convergence of the Taylor expansion.
 A possible solution is offered by a recent study~\cite{Morabit:2024sms}, which shows that additional diagrams not included in the current description could contribute to the \bbvv decay.
 As a result, the $\xi_{51}$ extracted from the fit would actually represent an effective value, to be rescaled by a (yet unknown) factor $<1$.
 At the same time, the complex calculations of the NMEs might miss some unaccounted cancellation of contributions from the intermediate nucleus in the $M^{2\nu}_{\rm GT-3}$ transition, which might lead to small values of $\xi_{31}$ beign compensated by larger-than-expected values of $\xi_{51}$.
 Finally, other effects not currently included, such as higher partial waves or weak magnetism~\cite{Morabit:2024sms}, might also play a role; so could be Beyond the Standard Model (BSM) physics~\cite{Bossio_2024}.
 From the combined marginalized posterior distributions of $\xi_{31}$ and $\xi_{51}$, we can rule out the HSD at more than $5\sigma$.

 \medskip
 In this Letter, we reported the most precise measurement of the \ce{^{130}Te} \bbvv \mbox{half-life} to date.
 The construction of a complete background model of \CUORE, together with the optimization of the data selection, translated into a twofold improvement with respect to our previous measurement.
 We also presented the first study of the \bbvv shape for \ce{^{130}Te} within the improved formalism.
 Here, the higher signal-to-background together with the background model, allowed us to obtain complementary constraints for the effective value of \gA with similar searches performed with other isotopes. We confirmed a preference for the SSD over the HSD model in the description of the \bbvv process, the latter being excluded at more than $5\sigma$.
 We found that, while $\xi_{31}$ meets the theoretical predictions, the value of $\xi_{51}$ is found to be far from the expectations.
 The discrepancy might arise from an incomplete theoretical description of the decay, such as minor effects not yet included or potential BSM physics. This motivates further theoretical studies of this transition and calls for a more precise experimental determination of the \bbvv spectral shape for \ce{^{130}Te}.

\begin{acknowledgments}
 The CUORE Collaboration thanks the directors and staff of the Laboratori Nazionali del Gran Sasso and the technical staff of our laboratories. This work was supported by the Istituto Nazionale di Fisica Nucleare (INFN); the National Science Foundation under Grant Nos. NSF-PHY-0605119, NSF-PHY-0500337, NSF-PHY-0855314, NSF-PHY-0902171, NSF-PHY-0969852, NSF-PHY-1307204, NSF-PHY-1314881, NSF-PHY-1401832, and NSF-PHY-1913374; Yale University, Johns Hopkins University, and University of Pittsburgh. This material is also based upon work supported by the US Department of Energy (DOE) Office of Science under Contract Nos. DE-AC02-05CH11231 and DE-AC52-07NA27344; by the DOE Office of Science, Office of Nuclear Physics under Contract Nos. DE-FG02-08ER41551, DE-FG03-00ER41138, DE- SC0012654, DE-SC0020423, DE-SC0019316. This research used resources of the National Energy Research Scientific Computing Center (NERSC). This work makes use of both the DIANA data analysis and APOLLO data acquisition software packages, which were developed by the CUORICINO, CUORE, LUCIFER, and CUPID-0 Collaborations. The authors acknowledge Advanced Research Computing at Virginia Tech for providing computational resources and technical support that have contributed to the results reported within this paper.
 This work is also financially supported by MCIN/AEI/10.13039/501100011033 from the grants: PID2023-147112NB-C22; CNS2022-135716 funded by the ``European Union NextGenerationEU/PRTR''; and CEX2019-000918-M to the ``Unit of Excellence Mar\'ia de Maeztu 2020-2023'' award to the Institute of Cosmos Sciences. F. \v{S}imkovic acknowledges support from the Slovak Research and Development Agency under Contract No.\ APVV-22-0413, Czech Science Foundation (GA\v{C}R), project No.\ 24-10180S.
 O. Ni\c{t}escu acknowledges support from the Romanian Ministry of Research, Innovation, and Digitalization through Project No. PN23 21 01 01/2023.
 O. Ni\c{t}escu  and J. Kotila acknowledge support from the Romanian Ministry of Research, Innovation, and Digitalization through Project PNRR-I8/C9-CF264, Contract No. 760100/23.05.2023 (The NEPTUN project).
\end{acknowledgments}

\bibliographystyle{apsrev4-1}
\bibliography{ref}

\end{document}